\documentclass{appolb}
\usepackage{graphicx}
\usepackage{amsmath}

\begin{document}
\title{X-rays help to unfuzzy the concept of measurement 
}

\maketitle


\author{C. Curceanu$^a$$^b$, B. C. Hiesmayr$^c$, K. Piscicchia$^a$$^b$
\newline

{$^a$}INFN, Laboratori Nazionali di Frascati, CP 13, Via E. Fermi 40, I-00044, Frascati (Roma), Italy\\
{$^b$}Museo Storico della Fisica e Centro Studi e Ricerche ``Enrico Fermi'', – Compendio del Viminale  –  Piazza del Viminale 1, 00184 Roma, Italy\\
{$^c$}University of Vienna, Faculty of Physics, Boltzmanngasse 5, 1090 Vienna, Austria\\
}

\begin{abstract}

In the last decades huge theoretical effort was devoted to the development of consistent theoretical models, aiming to solve the so-called ``measurement problem'', to which John Bell dedicated part of his thoughts. Among these, the Dynamical Reduction Models possess the unique characteristic to be experimentally testable, thus enabling to set experimental upper bounds on the reduction rate parameter $\lambda$ characterizing these models. Analysing the X-ray spectrum emitted by an isolated slab of Germanium, we set the most stringent limit on the parameter $\lambda$ up to date.

\end{abstract}

\section{Introduction}

One of John Bell's main concern was related to the measurement concept. On this he wrote ``\emph{The concept of measurement becomes so fuzzy on reflection that it is quite surprising to have it appearing in physical theory at the most fundamental level... does not any analysis of measurement require concepts more fundamental than measurement? And should not the fundamental theory be about these more fundamental concepts?}'' (Quantum Mechanics for Cosmologists (1981)). Indeed, the results of the measurements, related to the collapse of the wave function, generated the so-called ``measurement problem'', which continues to trigger intensive debates \cite{bou}. 

Among the proposed possible solutions to the ``measurement problem", the collapse models occupy a special place, since they are proven to be mathematically consistent models, so far in agreement with all known experimental predictions. On the top of it, these models make precise predictions and the aim of this contribution is to present a physical system that has the potential to test these predictions experimentally. 
In general, collapse models provide a theoretical framework for understanding how classical world emerges from quantum mechanics. Their dynamics practically preserves quantum linearity for the microscopic systems, but becomes strongly nonlinear when moving towards macroscopic scale. 

The conventional approach to test collapse models is to generate spatial superpositions of mesoscopic systems and examine the loss of interference, while environmental noises are under control. Naturally oscillating systems create quantum superpositions and thus represent a natural case-study for testing quantum linearity. Neutrinos, neutral mesons and chiral molecules are such systems. However, the collapse models can not be tested with neutrinos and the effect, stronger for neutral mesons, is still beyond experimental reach, while chiral molecules can offer promising candidates for testing collapse models \cite{bahrami}.

The best testing ground is offered by the spontaneous emission of radiation, predicted by the collapse models, used to settle the most stringent limit on the collapse models' characterizing parameter $\lambda$.

This paper is organized by introducing the physics behind collapse models (Sect. \ref{dynamical}), followed by presenting the computation of the spontaneous radiation emission rate of free electrons (Sect. \ref{spontaneous}). Then we present our new limit on the reduction rate parameter $\lambda$, and conclude in Sect. \ref{conclusions}.

\section{Dynamical Reduction Models: the wave function localization mean rate parameter $\lambda$}\label{dynamical}

The recent development of experimentally testable, mathematically consistent Dynamical Reduction Models, as a possible solution of the so-called {\em measurement problem} strongly renewed the interest of the scientific community for the foundations of Quantum Mechanics (QM). 

Dynamical Reduction Models were born to answer the long standing problem of measurement in QM and to settle in a more natural way its role as a theory of both micro and macroscopic phenomena. A scheme for an ideal measurement process was already used by von Neumann \cite{neumann} to show that the linear nature of the Schr\"odinger equation enables for the superposition of macro-object states (a general demonstration was more recently obtained by Bassi and Ghirardi \cite{bassi}). Under the assumption that QM is a complete theory only two ways out exist:
\begin{itemize}
\item the existence of two dynamical regimes: a) the quantum states evolution governed by the Schr\"odinger equation, unitary and linear and, thus, deterministic b) the measurement process, governed by the wave packet reduction principle, which is non-linear and intrinsically stochastic,
\item  a non-linear stochastic modification of the Hamiltonian dynamics. 
\end{itemize}

The first consistent and satisfying  Dynamical Reduction Model, known as Quantum Mechanics with Spontaneous Localization (QMSL), was developed by Ghirardi, Rimini and Weber \cite{ghi}. According to the QMSL model, particles undergo spontaneous localizations around definite positions, following a Possion distribution characterized by a  mean frequency   $\lambda$ that was set to  $10^{-16}$ s$^{-1}$. During 1989-90 the efforts of Ghirardi, Rimini, Weber and Pearle \cite{pear} led to the development of the CSL model (Continuous Spontaneous Localization). The CSL theory is based on the introduction of new, non linear and stochastic terms in the Schr\"odinger equation. Such terms induce, for the state vector, a diffusion process, which is responsible for the wave packet reduction (for reviews and references see \cite{review}).  
The Dynamical Reduction Models posses the unique characteristic to be experimentally testable, by measuring the small predicted deviations with respect to the standard QM. Q. Fu \cite{fu} demonstrated that the particle interaction with the stochastic field, besides collapsing the state vector on the position basis, causes an enhancement of the energy expectation value. This implies, for a charged particle, the emission of electromagnetic radiation (known as spontaneous radiation)  not present in the standard QM (see for illustration also Figure \ref{electron}). A measurement of the emitted radiation rate thus enables to set a limit on the $\lambda$ parameter of these models.

By comparing the expected emission rate with the radiation emitted by an isolated slab of Germanium at 11 KeV \cite{miley}, Fu obtained an upper limit on the reduction rate parameter ($\lambda < 0.55 \cdot 10^{-16} s^{-1}$). In the next sections the arguments of Fu will be described and  a more refined analysis of the X-ray emission spectrum measured by the IGEX experiment \cite{igex1,vip} will be presented, which improves the precedent limit by a factor 20.

\section{The spontaneous emission rate}\label{spontaneous}

The radiation spectrum spontaneously emitted by a free electron (see Figure \ref{electron} for a schematic representation), as a consequence of the interaction with the stochastic field, was calculated by Fu \cite{fu} in the framework of the non-relativistic CSL model, and it is given by:

\begin{figure}[htb]
\centerline{%
\includegraphics*[height=4cm,width=11cm]{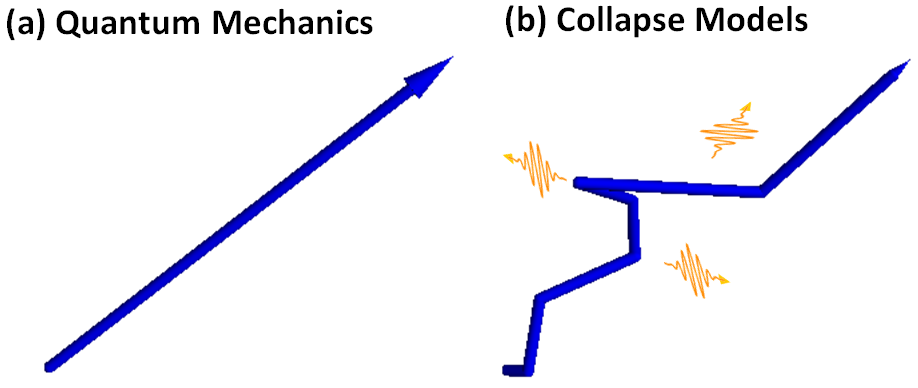}}
\caption{\em
Schematic representation of the spontaneous emission process caused by the interaction of a free charged particle with the stochastic field.}
\label{electron}
\end{figure}

\begin{equation}\label{furate}
\frac{d\Gamma (E)}{dE} = \frac{e^2 \lambda}{4\pi^2 a^2 m^2 E}
\end{equation}
where $m$ represents the electron mass, $E$ is the energy of the emitted photon, $\lambda$ and $a$ are respectively the reduction rate parameter and the correlation length of the reduction model (the latter is generally accepted to be of the order of $a=10^{-7}$m). Note that both parameters $\lambda$ and $a$ would play the role of natural constants if the models turn out to be correct. 

If the stochastic field is assumed to be coupled to the particle mass density (mass proportional CSL model) (see for example \cite{bassi}) then the previous expression for the emission rate eqn. (\ref{furate}) for electrons is to be multiplied by the factor $(m_e/m_N)^2$, with $m_N$ the nucleon mass. Using the measured radiation appearing in an isolated slab of Germanium \cite{miley} corresponding to an energy of 11 KeV, and employing the predicted rate eqn. (\ref{furate}), Fu obtained the following upper limit for $\lambda$:

\begin{equation}\label{fulimit}
\lambda < 0.55 \cdot 10^{-16} s^{-1}.
\end{equation}
Only the four valence electrons were considered to contribute to the measured X-ray emission; since their binding energy $\sim 10$ eV is orders of magnitude less than the emitted photons' energy, they can be considered as \emph{quasi-free}.

In Ref. \cite{Adler} the author argues that, in evaluating his numerical result, Fu uses for the electron charge the value $e^2=17137.04$, whereas the standard adopted Feynman rules require the identification $e^2/(4\pi)=17137.04$. We took into account this correction when evaluating the new limit on the collapse rate parameter presented in Section \ref{fitpar}.

\section{Determination of a new limit on $\lambda$}

In order to reduce possible biases introduced on the $\lambda$ value by the punctual evaluation of the rate at one single energy bin, we decided to adopt a different strategy with respect to the analysis performed in \cite{fu}. The X-ray emission spectrum measured by the IGEX experiment \cite{igex1} was fitted in the range $\Delta E =$ $4.5\div 48.5$ KeV $\ll m$, compatible with the non-relativistic assumption (for electrons) used in the calculation of the predicted rate (eqn. (\ref{furate})). IGEX is a low-background experiment based on low-activity Germanium detectors dedicated to the neutrinoless double beta decay ($\beta \beta 0 \nu$) decay research. The published data \cite{igex2}, used to extract a new upper limit on the collapse rate parameter, refers to 80 kg day exposure.

A Bayesian model is adopted to calculate the $\chi^2$ variable minimized to fit the X-ray spectrum, assuming the predicted (eqn. (\ref{furate})) energy dependence:

\begin{equation}\label{fitfunc}
\frac{d\Gamma (E)}{dE} = \frac{\alpha(\lambda)}{E}.
\end{equation}

\subsection{Fit result and discussion}\label{fitpar}

\begin{figure}[htb]
\centerline{%
\includegraphics*[height=9cm,width=9cm]{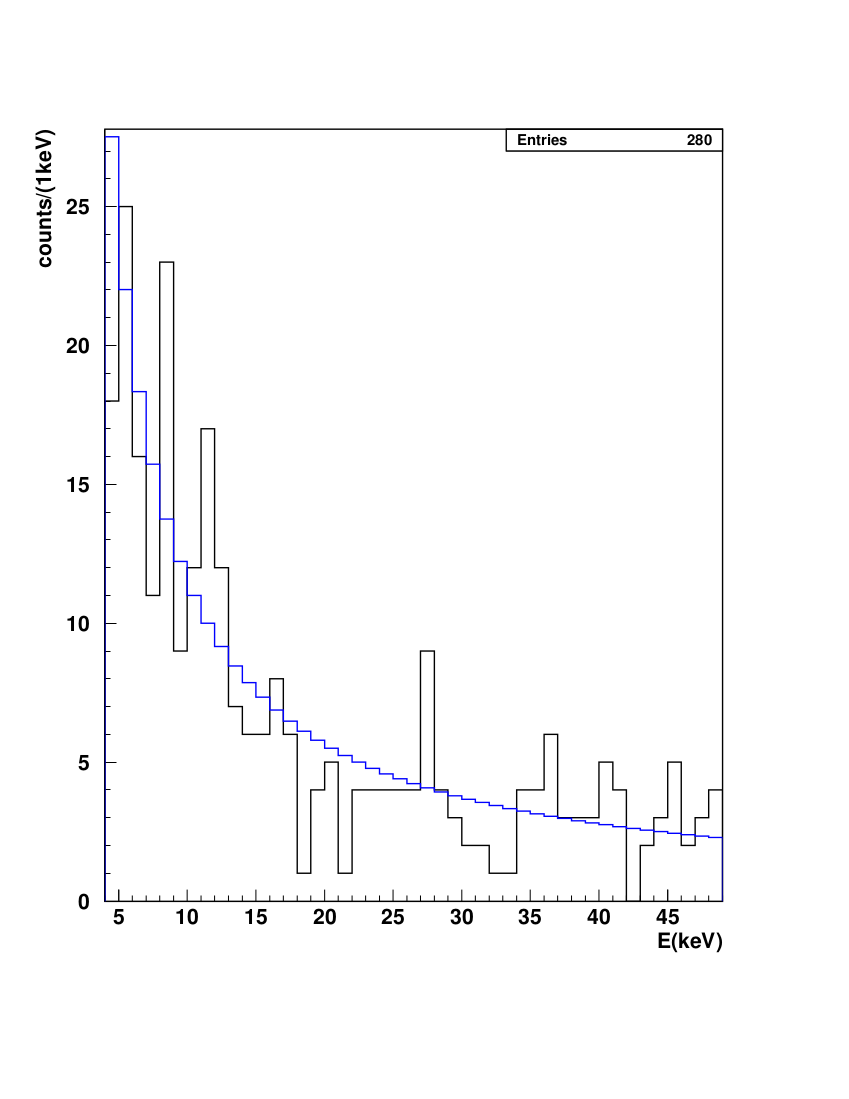}}
\caption{\em
Fit of the X ray emission spectrum measured by the IGEX experiment \cite{igex1,igex2}, using the theoretical fit function eqn. (\ref{fitfunc}).}
\label{fit}
\end{figure}

The result of the performed fit is shown in Figure \ref{fit}. The minimization gives for the free parameter of the fit the value $\alpha(\lambda) = 110 \pm 7$, corresponding to a reduced chi-square $\chi^2/n.d.f = 1.1$.

An upper limit on the parameter $\lambda$ can then be set according to the result of eqn. (\ref{furate}):

\begin{equation}\label{fu1}
\frac{d\Gamma (E)}{dE} = c \, \frac{e^2 \lambda}{4\pi^2 a^2 m^2 E}  \leq \frac{110}{E},
\end{equation}
where the factor $c$ is given by:

\begin{equation}
c = (8.9 \,\, 10^{24})\,\,(8.6\,\, 10^4)\,\,(4),
\end{equation}
the first bracket accounts for the particle density of Germanium, the second term is the number of seconds in one day and 4 represents the number of valence electrons in Germanium. Considering as quasi-free the four valence electrons, consistently with  Fu's hypothesis, applying eqn. (\ref{fu1}) and using the correct prescription $e^2/(4\pi)=17137.04$ (see \cite{Adler}), the following upper limits for the reduction rate parameter is obtained:

\begin{equation}\label{lim2}
\lambda \leq 1.4 \cdot  10^{-17} s^{-1} \;\;\;\mbox{non mass prop.} \;\; ; \;\; \lambda \leq 4.7 \cdot  10^{-11} s^{-1} \;\;\;\mbox{mass prop.}
\end{equation}
if the stochastic field is not assumed to be coupled to the particle mass density (left) and for a mass proportional CSL model (right).
If we consider in the calculation the 22 external electrons of the ${}^{32}Ge$ atom, down to the $3s$ orbit, based on the fact that the corresponding binding energy ($BE_{3s}=180.1$ eV) is about 22 times smaller than the less energetic measured photons (the 22 outermost electrons can be then considered to be \emph{quasi-free}) the limits on the reduction rate parameter become:

\begin{equation}\label{lim4}
\lambda \leq 2.5 \cdot  10^{-18} s^{-1} \;\;\;\mbox{non mass prop.}
\end{equation} 
\begin{equation}\label{lim5}
\lambda \leq 8.5 \cdot  10^{-12} s^{-1} \;\;\;\mbox{mass prop.}
\end{equation} 
The limits in eqns. (\ref{lim4} and \ref{lim5}) improves the precedent Fu's limit, eqn. (\ref{fulimit}), by a factor 20. The estimated upper limits eqns. (\ref{lim4} and \ref{lim5}) for the rate parameter are to be compared with the values conventionally assumed in the specific collapse models:

\begin{equation}\label{lmodel}
\lambda_{QMSL} = 10^{-16} s^{-1} \qquad , \qquad   \lambda_{CSL} = 2.2 \cdot  10^{-17} s^{-1}
\end{equation}
and with the values proposed, more recently, by S. Adler \cite{Adler}. From eqn. (\ref{lim4}) one concludes that if a mass proportional coupling is not taken into account, and by considering a ``white'' noise inducing the collapse, the upper limit on $\lambda$ strongly constrains the collapse models, being exceeded by both $\lambda_{QMSL}$ and $\lambda_{CSL}$. 

In Table \ref{upper} the upper bounds on the $\lambda$ parameter from laboratory experiments (updated with the present result in the mass proportional assumption) and astronomical observations are summarized.

\section{Conclusions and perspectives}\label{conclusions}

The collapse of the wave function and, more generally the ``measurement problem'' is one of the hottest topics in QM and was attracting John Bell's attention. A possible mechanism inducing the collapse, the so called Continuous Spontaneous Localization (CSL), has an unique experimental signature: a spontaneous radiation emitted by (free) charged particles. A new limit on the free collapse frequency parameter $\lambda$, characterizing the CSL model, was obtained by performing an analysis of the IGEX experimental data. The $\lambda$ value is obtained to be  $\lambda \leq 2.5 \cdot  10^{-18} s^{-1}$ if no mass dependence is considered, and $\lambda \leq 8.5 \cdot  10^{-12} s^{-1}$ if such a dependence is taken into consideration, to be compared with proposed CSL value ($\lambda_{CSL} = 2.2 \cdot  10^{-17} s^{-1}$).

We are presently exploring the possibility to perform a new measurement
that will allow for more then 2-3 orders of magnitude improvement on the collapse rate parameter $\lambda$.

\begin{table}
\caption{\label{upper}Upper bounds on the $\lambda$ parameter ($\lambda_{CSL} = 2.2 \cdot  10^{-17} s^{-1}$). Numbers represent the distance (in orders of magnitude) from the CSL standard value.}
\begin{tabular}{llll}

&&& \\
Laboratory &  &Astronomical&  \\
experiments & &observations& \\
& && \\

Fullerene difraction &12-13&Dissociation of cosmic & 18\\
experiments&&hydrogen &\\

Decay of supercurrents &15&Heating of Intergalactic&9\\
 (SQUIDS)&&medium (IGM)&\\

Spontaneous X-ray &5&Heating of protons in&13\\
emission from Ge&& the universe&\\
Proton decay&19&Heating of Interstellar &16\\
&&dust grains&\\

\end{tabular}

\end{table}

\section*{Acknowledgments}

The support from the HadronPhysics FP6(506078), HadronPhysics2 FP7 (227431), HadronPhysics3
(283286) projects and from the Austrian Science Foundation FWF-P26783 project is gratefully acknowledged. We acknowledge as well for
 the support of the EU COST Action MP1006, \emph{Fundamental Problems in Quantum Physics}, and
from the Museo Storico della Fisica e Centro Studi e Ricerche ``Enrico Fermi" (”\emph{Problemi aperti nella meccania quantistica” project}).

\end{document}